\documentclass[conference]{IEEEtran}
\IEEEoverridecommandlockouts
\usepackage{cite}
\usepackage{amsmath,amssymb,amsfonts}
\usepackage{algorithmic}
\usepackage{graphicx}
\usepackage{textcomp}
\usepackage{xcolor}
\usepackage{mathtools}
\usepackage{amsmath}
\def\BibTeX{{\rm B\kern-.05em{\sc i\kern-.025em b}\kern-.08em
    T\kern-.1667em\lower.7ex\hbox{E}\kern-.125emX}}
\begin{document}

\title{Sparse Reconstruction of Chirplets for Automotive FMCW Radar Interference Mitigation\\
{\footnotesize }
}

\author{\IEEEauthorblockN{1\textsuperscript{st} Aitor Correas-Serrano}
\IEEEauthorblockA{\textit{Cognitive Radar} \\
{Fraunhofer FHR}\\
Wachtberg, Germany \\
aitor.correas@fhr.fraunhofer.de}
\and
\IEEEauthorblockN{2\textsuperscript{nd} Mar\'{i}a A. Gonz\'{a}lez-Huici}
\IEEEauthorblockA{\textit{Cognitive Radar} \\
\textit{Fraunhofer FHR}\\
Wachtberg, Germany \\
maria.gonzalez@fhr.fraunhofer.de}
}

\maketitle

\begin{abstract}
Mutual interference in automotive radar scenarios is going to become a major concern as the density of vehicles with radar sensors in the roads increases. The present work tackles the problem of frequency modulated continuous wave (FMCW) to FMCW and continuous wave interference. In this context, we propose a signal processing technique to blindly identify and remove interference by using the fast Orthogonal Matching Pursuit (OMP) algorithm to project the interference signals in a reduced chirplet basis, and separate it from the target signal with minimal loss of information. Significant reduction of the noise-plus-interference levels are observed in both simulated and measured data, the later acquired with state of the art automotive sensors. 
\end{abstract}


\section{Introduction}

Radar is a key enabling technology for advanced driving assistant systems (ADAS) and autonomous driving due to its ability to detect motion and localize targets at long distances regardless of the weather and lighting conditions. The rise of popularity of this technology in the automotive field, further motivated by the advances in the production of low-cost radar chips, will make radar sensors ubiquitous in vehicles with high levels of automation. As increasing amount of these vehicles share the roads, car-to-car interference will become an issue that needs to be dealt with. Car-to-car radar interference, if not mitigated or avoided, degrades the detection capabilities of the sensors, effectively disabling them in dense traffic scenarios with high interference \cite{IntStudy}\cite{IntStudy2}\cite{IntStudy3}.

State of the art automotive radars use FMCW modulations restricted to the band of 76 GHz to 81 GHz, with short range radars (SRR) occupying most of the available bandwidth. In these conditions, interference is bound to happen. After the interference is identified, it could be avoided with frequency hopping techniques, as shown in \cite{freqHopp}. In crowded environments, however, there is no guarantee that the new band will be interference free, in particular given that SRR radars may use most of the available  spectrum. Other methods try to reconstruct the original signal assuming  an approximate knowledge of the interference time \cite{CancelInt}, in some cases removing the interfered portions and performing range estimation with the remaining signal \cite{InterfSparse}. The removal of the interfered samples not only assumes knowledge of the time of interference, but it is also impractical for scenarios in which the majority of the signal is corrupted, as the remaining fragmented signal might not be enough to achieve a good estimation. More recent work blindly separates the interference from the target signal by solving a dual basis pursuit problem in the short-time Fourier transform (STFT) and standard Fourier basis \cite{Faruk}. However, the STFT basis is not ideal for representing fast wide-band events such as FMCW interference, due to its inherent trade-off between frequency and time resolution. The proposed method in \cite{Faruk} also requires a joint estimation of target and interference signals, reducing the flexibility of the approach. Lastly, basis pursuit tends to be more computationally complex and less suitable  for hardware acceleration when compared to matching pursuit approaches \cite{CScomparison}\cite{CScomparison2}\cite{FPGAOMP}. 

\begin{figure}[t]
  \centering
  \includegraphics[width=1\linewidth]{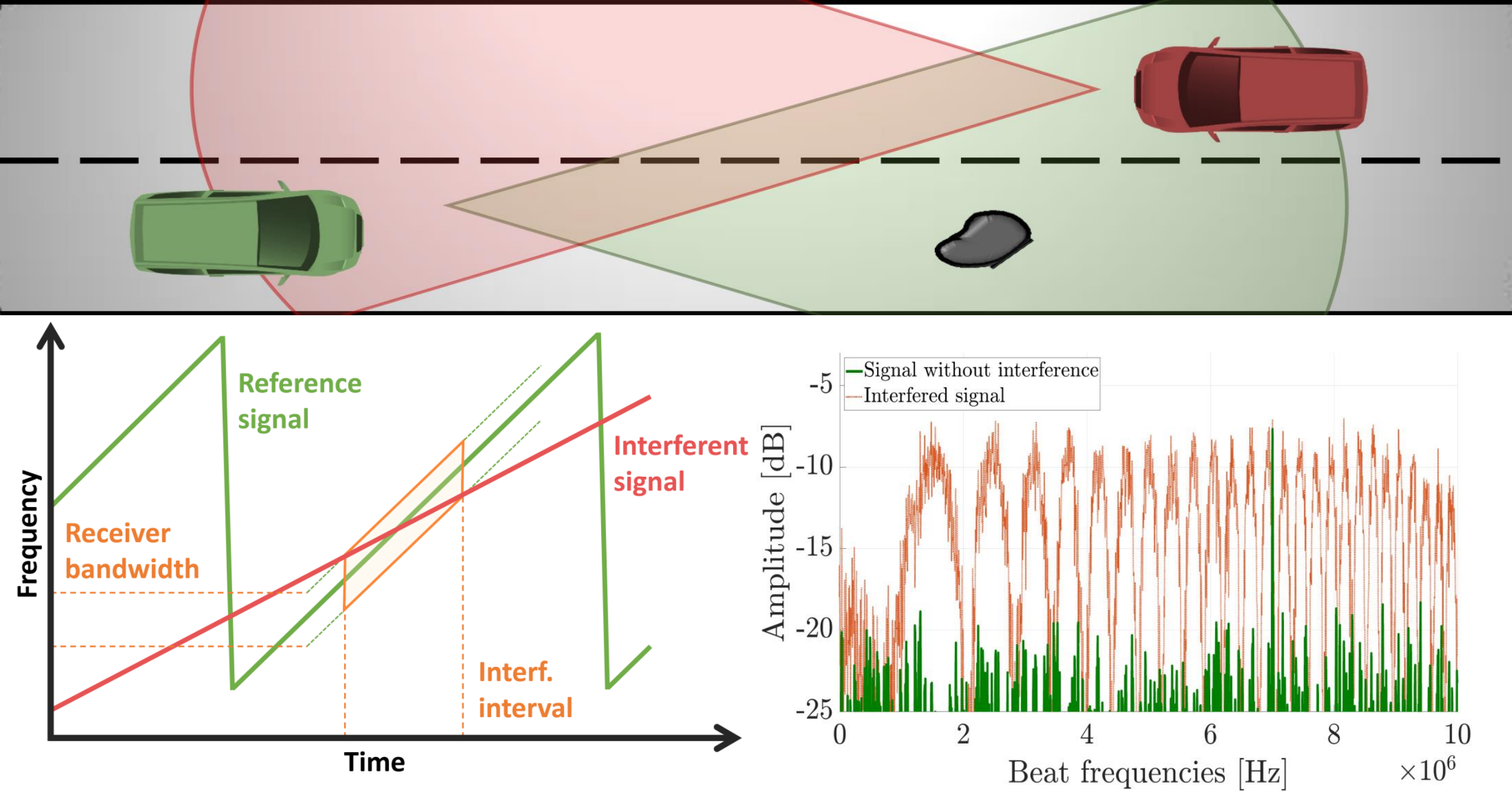}
  \caption[]{Interference can potentially impede the detection capabilities of the radar sensors, creating dangerous situations.}
\label{fig:Arrays}
\end{figure}

In FMCW processing, targets take the form of tones in baseband, whereas external FMCW interference with different modulation rate appears as a chirp. The chirp-like shape of the interference can be exploited to blindly detect and eliminate the interference by reconstructing the signal on a reduced chirplet transform basis. The chirplet transform decomposes the signal in time-windowed chirps \cite{chirpletTransform} defined by their time delay, duration, starting frequency and modulation rate. In the case of FMCW radar interference, the bandwidth is set by the anti-aliasing filter in the receiver. This results in a reduced basis in which only the time delay and the duration of the interference remain as unknown parameters, lowering the computational complexity of the transform.

In this paper, we present a method to mitigate the interference in FMCW automotive radar by reconstructing the interference in the reduced chirplet basis using the Orthogonal Matching Pursuit (OMP) \cite{OMPexplained} algorithm. In Section II we briefly explain the FMCW radar model in the presence of interference, illustrating how interference appears in the receiver after down-conversion and filtering. Subsequently, we justify the use of a reduced chirplet transform as basis for the interference, and propose a fast method to mitigate it applying OMP sparse reconstruction with the reduced chirplet basis in Section III. Simulated results are presented and discussed in Section IV. Section V and Section VI show respectively the experimental setup for the data acquisition and the results of applying the proposed method in interfered automotive radar signals, showing significant improvements in the signal to noise-plus-interference ratio (SNIR). Section VII concludes the paper.

\section{FMCW signal model with interference}

In this section we shortly describe the signal model for FMCW radar in the presence of FMCW interference. Automotive radar sensors operate generally with real-valued samples, hence we work with the sine function in time domain rather than complex exponential terms. The most common waveform in automotive FMCW radar is the linearly frequency modulated signal (chirp), defined as 
\begin{equation}\label{eqn:reference}
s_{ref} = \sin(2\pi (f_c t + \frac{k}{2} t^2)) \quad \textrm{    for    }  \quad  0 < t < T
\end{equation}
with T being the chirp duration, k being the modulation rate defined as the \(B/T\), and B being the bandwidth. This signal also serves as reference signal for the down-conversion process in the receiver. The carrier frequency $f_c$ lays in the reserved automotive radar band, from 76 GHz to 81 GHz. 

If there are $N_i$ chirp-like interference signals from another FMCW radars operating in the same band with modulation rate $k_i$, and $N_t$ echoes from target reflections,  the signal at the receiver is the sum of the signal from the targets $s_R$ and the interference signals $s_I$: 
\begin{align}\label{eqn:signals}
s_R &= \sum_t^{N_t}P_t  \sin( 2 \pi (f_c (t -\tau_t) + \frac{k}{2} (t-\tau_t)^2)) \\
s_I &= \sum_i^{N_i}P_i  \sin( 2 \pi (f_c (t -\tau ) + \frac{k_i}{2} (t-\tau_i)^2)) 
\end{align}
where $\tau$ is the propagation delay, the $i$ sub-index corresponds to the parameters associated to the interference signals, and $t$ sub-index to the ones associated to the targets. $P_t$ stands for the amplitude of the target echo, whereas $P_i$ refers to the amplitude of the signal from the interfering radar. 

The received signal is converted to baseband through mixing with the complex conjugate (i.e., a phase delay of 180 degrees) of the reference signal \eqref{eqn:reference}
\begin{equation}\label{eqn:downconversion} 
y(t) = (s_R + s_I) s_{ref}^* = y_R + y_I 
\end{equation}
where 
\begin{align}\label{eqn:basebandsignals}
y_R &= \sum_t^{N_t}P_t \sin( \pi k (\tau_t^2 - 2 t \tau_t) + 2 \pi f_c \tau_t) \\
y_I &= \sum_i^{N_i}P_i \sin( \pi (k_i - k)t^2 - 2 k_i t \tau_i + k_i \tau_i^2 + 2 \pi f_c \tau). \label{eqn:basebandsignals2}
\end{align}

\begin{figure}[t]
  \centering
  \includegraphics[width=1\linewidth]{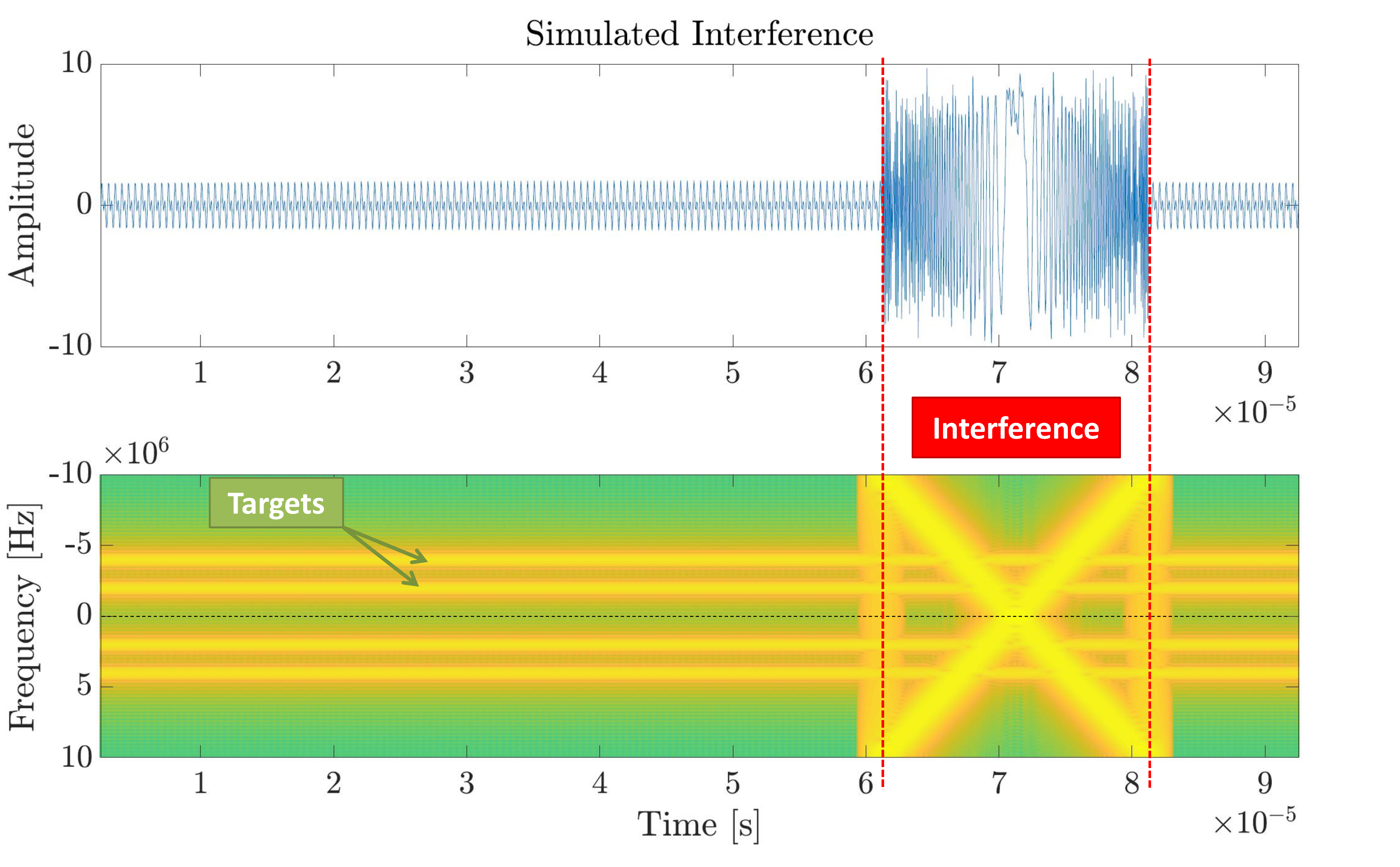}
  \caption[]{Simulated FMCW radar interference in baseband. The chirp-like time restricted nature of this type of interference is clear in the STFT representation of the signal. This transform, however, is not adequate for interference mitigation due to the trade-off between time and frequency accuracy inherent to the STFT basis.}
\label{fig:IntSim}
\end{figure}

The first term of the summation in \eqref{eqn:downconversion} corresponds to a target echo, appearing as a tone at a particular beat frequency. The second term accounts for the interference, with a quadratic phase term that corresponds to a linearly modulated frequency (a chirp) in baseband. After down-conversion, the bandwidth of the signal is limited by an analog low-pass filter, with a pass frequency $f_r = f_s/2$, in order to avoid aliasing. In further derivations, we consider a perfect low-pass filter, so no components over the sampling frequency appear in the signal. The instantaneous frequency of the interference is, as a result, bounded by the receiver
\begin{equation}\label{eqn:flimit}
|f_i(t)| =  |(k_i - k) t - 2 k_i \tau^2| \leq f_r,
\end{equation}
as is the duration
\begin{equation}\label{eqn:tlimit}
T_i \leq  \left| \frac{2 f_{r}}{(k_i - k)}\right|.
\end{equation}

After establishing the time and frequency bounds of the interference signal through the receiver bandwidth, we can finally write the received signal in baseband as

\begin{equation}
     \label{eq:partbbqe}
     y(t) = \left\{
	       \begin{array}{ll}
		 y_R + y_I \quad    \mathrm{if} \enskip  \frac{k_i \tau_i - f_r}{k_i - k} \le t \le \frac{k_i \tau_i + f_r}{k_i - k} \\
		 y_R      \quad \quad \quad       \mathrm{otherwise}.
	       \end{array}
	     \right.
\end{equation}

 Fig. \ref{fig:IntSim} shows a simulated scenario in which we can observe interference in both time domain as well as the corresponding short time Fourier transform (STFT) domain. The signal is even in frequency, as we consider real-valued samples. The chirp-like, time limited nature of the interference becomes clear in the STFT representation.

\section{Method for Interference identification and mitigation}

The proposed method relies on a projection of the received signal in a reduced chirplet transform basis. To find this projection, we use the Orthogonal Matching Pursuit (OMP) \cite{OMPexplained} algorithm, as it is fast and it can be further accelerated through embedded programming, making it a great choice to achieve real time computation with actual automotive radars.

\subsection{Reduced chirplet transform basis}

The chirplet transform reconstructs a signal as a summation of time limited chirps of varying characteristics and amplitude. Traditionally, the chirplet transform is a 4-parameter transform. The atoms of the chirplet basis are
\begin{equation}\label{eqn:chirplet}
c_{\nu,\tau,\delta,\kappa}(t) = \sin(2\pi (f_\nu t + \frac{k_\kappa}{2} t^2)) \quad  T_\tau < t < (T_\tau + T_\delta),
\end{equation}
with the four parameters characterized in the sub-indexes $[\nu,\tau,\delta,\kappa]$ representing the different possible values of frequency-shift, time-shift, duration and modulation-rate of the chirp hypothesis. Such a transform is costly to compute. In the context of FMCW interference, however, we can define a reduced chirplet transform in which some of these parameters are defined by the characteristic of the receiver. Namely, the frequency shift is set to be the cut-off frequency of the low pass filter $f_r$, whereas slope and duration turn into coupled parameters also defined by the bandwidth of the receiver, as per \eqref{eqn:tlimit}, eliminating the need to estimate one of them. As a result, we define a new, reduced chirplet basis, whose chirplets are defined as
\begin{equation}\label{eqn:restchirplet}
c_{\tau,\kappa}(t) = \sin(2\pi (f_r t + \frac{k_\kappa}{2} t^2)) \quad  T_\tau < t < (T_\tau + T_i),
\end{equation}

Selecting $M_{\kappa}$ hypothesis for the slopes and $M_{\tau}$ hypotheses of the time-shift, we can create a dictionary for our reduced chirplet transform 
\begin{equation}\label{eqn:Dictionary}
A = [c_{11}, ..., c_{\tau 1}, ..., c_{1 \kappa}, ..., c_{\tau \kappa}] \in \mathbb{R}^{N \times M}.
\end{equation}
with $M = M_{\kappa}M_{\tau}$. For the reconstruction of the interference, we choose values of $|k_\kappa| > 0$, as we do not want to include the actual targets in our dictionary.

\subsection{Orthogonal Matching Pursuit for interference mitigation}

OMP is well known for its use in sparse reconstruction problems in compressed sensing, and has seen some use in the radar field. It is a fast greedy algorithm suitable for hardware accelerated implementations \cite{HardAccel1} \cite{HardAccel2}, making it ideal for automotive radar, where cost and computational resources are main limiting factors.

OMP finds a solution to $y = Ax$ in $x$, where $y \in \mathbb{R}^N$ is the measurement (the samples corresponding to a chirp), $A$ is our chirplet basis and $x \in \mathbb{C}^M$ is the interference reconstruction in the chirplet basis. For an M-long grid of hypotheses, OMP first estimates hypothesis $m$ of the dictionary $A$ with the maximum correlation with the current residual
\begin{equation}\label{eqn:match response}
\textrm{argmax}_i \quad m := A^H r_{it-1}, 
\end{equation}
where $i$ is the hypothesis index, and $it$ is the iteration counter, with $r_{it}$ being the residual signal, initialized as $r_0 = y$. The grid point $m$ is stored in a set $I^{it} = I^{it-1}\cup m$. The columns of $A$ listed in $I^{it}$  are selected to form the partial sensing matrix $E$, through which the complex coefficients for the selected hypothesis are calculated by finding $min_w ||y - E w||_2$. The residual is updated as
\begin{equation}\label{eqn:residue}
r_{it} = y - E w,
\end{equation} and the algorithm continues to the next iteration. The algorithm stops once $r_{it}$ falls below a given threshold, or after a predefined number of iterations. The reconstructed signal $x$ is obtained by setting the entries indexed in $I^{it}$ to the corresponding value of $w_{it}$, leaving the rest of entries of $x$ at zero.

By setting $A$ as our reduced chirplet basis, we can see how we obtain information about the interference based on the reconstructed $x$. More importantly, given a successful approximation of the interference, the residual after the application of the algorithm will have the interference components removed in \eqref{eqn:residue}, leaving only the target related components to apply any posterior processing. The Doppler and direction of arrival phase components in the target signals are preserved. Note that OMP could be used to estimate also the targets using a fitting model \^A with beat frequency tones as hypotheses. We choose to use OMP to reconstruct only the interference (i.e. not including target echo hypothesis in our model) with the intention of accurately showing the extent of the interference mitigation through the comparison of the spectral power density of the signals before and after the mitigation. 

\section{Simulations}

\begin{figure}[b]
  \centering
  \includegraphics[width=1\linewidth]{Figures/Interference_1Small_Asimetric.pdf}
  \caption[]{Interference mitigation in a simulated scenario with two targets and a strong interferer. We achieve an improvement of the SNIR to levels close to the non interfered signal.}
\label{fig:Interference_1Small}
\end{figure}

\begin{figure}[t]
  \centering
  \includegraphics[width=1\linewidth]{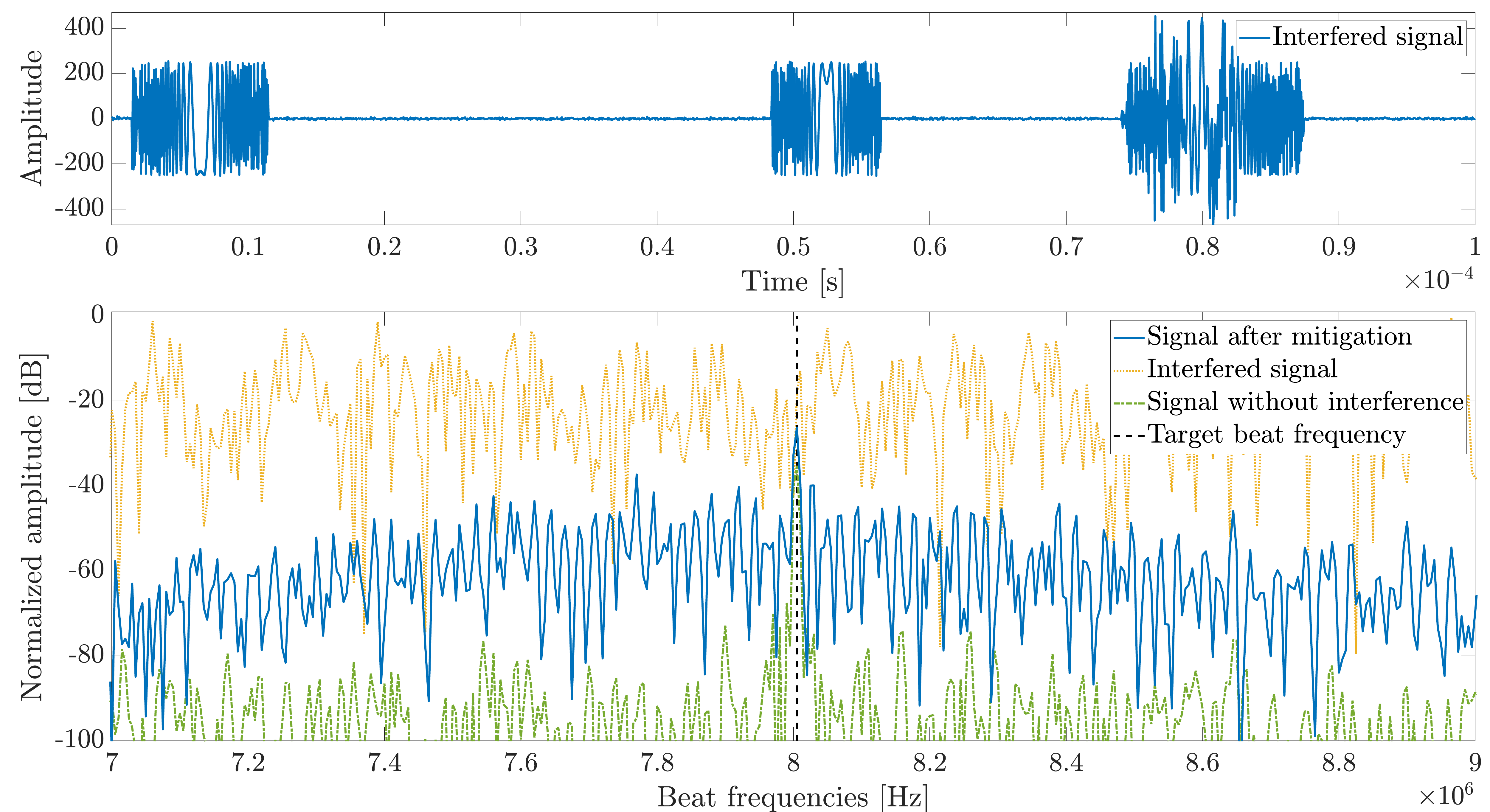}
  \caption[]{Interference mitigation in a simulated scenario with one target, four strong interferers, and interference overlapping. The improvement of 50 dB in the SNIR is good enough to discern the target from the noise-plus-interference.}
\label{fig:Interference_4}
\end{figure}

To test our approach we build a simulation environment in which we can generate interfered signals based on the presented model. Interference is a complicated event to measure in a controlled scenario, specially so in the case of multiple simultaneous interference. This simulation environment allows us to test our interference mitigation algorithm in a variety of hypothetical scenarios. We generate the interference off-grid, but we do not take into account the non-idealities in the receiver of a real sensor. Fig. \ref{fig:Interference_1Small} and Fig. \ref{fig:Interference_4} show the time and frequency domain of two synthetic received signals before and after the application of our interference mitigation algorithm. 

We apply a two step search using the proposed method: a first search over a coarse grid in the whole range of possible time-shifts and slopes, and a second search with a fine grid defined in the regions where interference is likely, according to the first search. For the first search, we use $M_{\kappa} = 200$ slope hypotheses and $M_{\tau} = 600$ time-shift hypotheses. In the second search we set $M_{\kappa} = 40$  and $M_{\tau} = 40$. The synthetic signal has $N = 2000$ samples in both cases. We use bigger values of $M_{\kappa}$  and $M_{\tau}$ in the first step as we search over a wider variety of time-delay and slope hypothesis. A measure of the variation of the energy in the residual term is used as stopping criteria for OMP.

Fig. \ref{fig:Interference_1Small} represents a relatively easy scenario, in which there is one strong interference that completely masks the target beat frequencies. After mitigation of the interference through the restricted chirplet transform we observe an estimation of similar quality to that of the synthetic signal without added interference. This represents an average improvement in the SNIR of over 35 dB. Fig. \ref{fig:Interference_4} depicts a more complex scenario with four interferers, each of them of higher power than the interferers in Fig. \ref{fig:Interference_1Small}, and with two of the interference signals overlapping in time. This scenario is particularly challenging due to the full overlapping between two independent interference signals and the high interference power, around 40 dB over the power of the echo from the target. The result of applying our interference mitigation algorithm is an increase of the SNIR level of an average of 50 dB, enough to distinguish the target over the remaining noise and non-mitigated interference.

While these case studies show the potential of our approach to mitigate interference in a variety of scenarios, there are a multitude of non-idealities occurring in real sensors that we are not accounting for. With this as motivation, we set up a simple measurement campaign to test our approach with actual measurements.

\section{Experimental setup}

\begin{table}[b]
\caption[]{FMCW waveform parameters of reference and interference}
  \centering
  \includegraphics[width=1\linewidth]{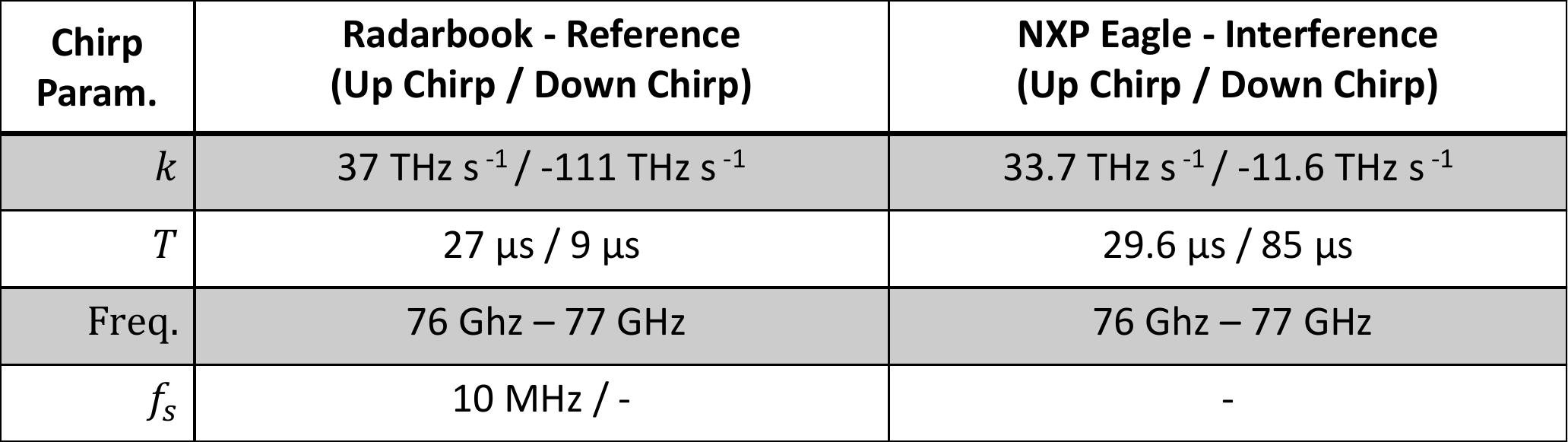}
\label{tab:Scenario}
\end{table}

\begin{figure}[t]
  \centering
  \includegraphics[width=1\linewidth]{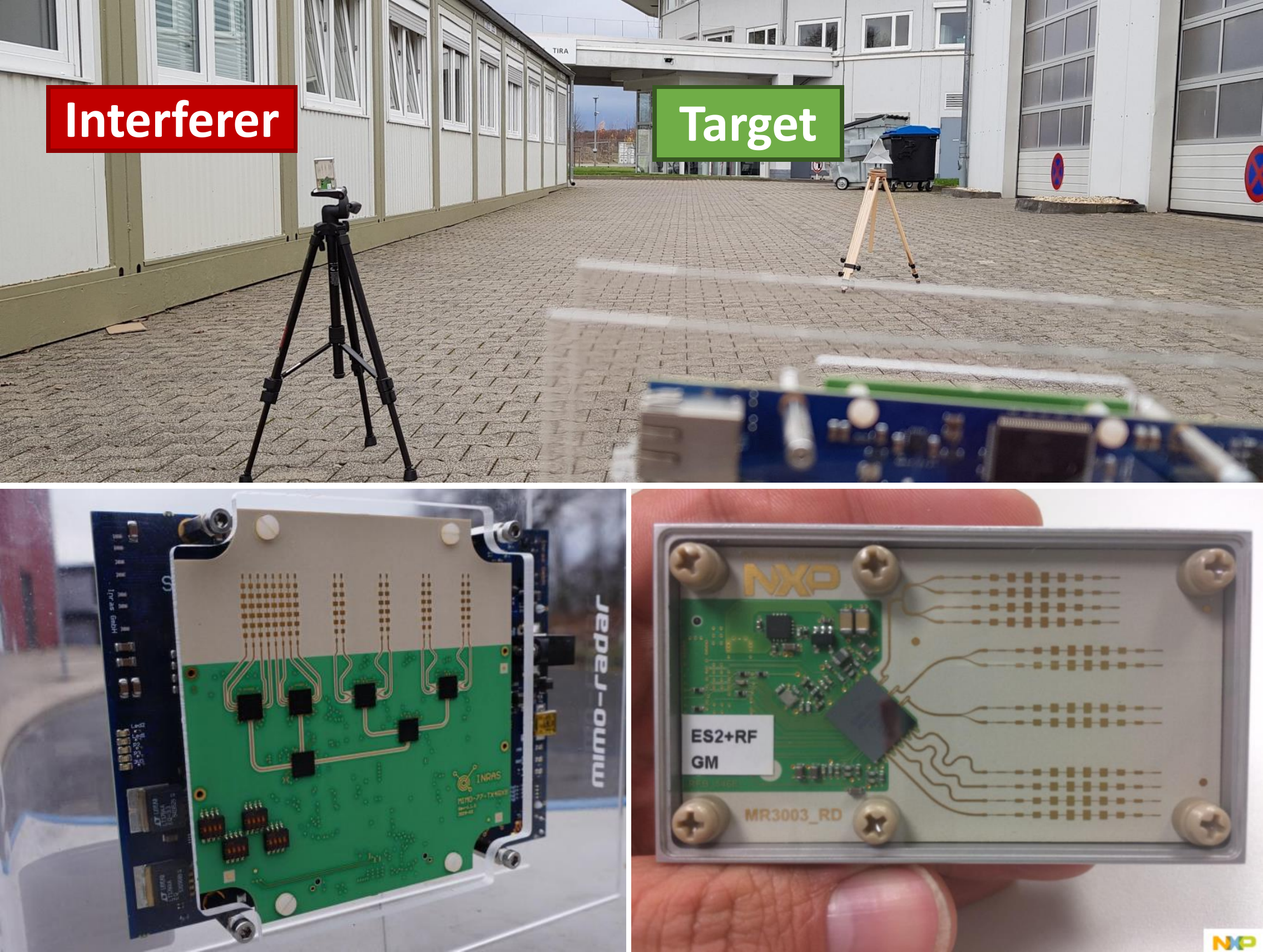}
  \caption[]{On the top, a picture of the measured scenario, with a target and an interference source. Radarbook (reference) on the bottom left and NXP Eagle (interference) on the bottom right.}
\label{fig:Scenario}
\end{figure}

To capture automotive radar interference, we set up an experiment in which we confront two different automotive radar sensors, simultaneously transmitting and receiving in the band of 76-77 GHz. The data used for the validation of our algorithm is captured by the 4Tx/8Rx Radarbook system, commercialized by Inras \cite{RBK}. The interfering sensor is a prototype NXP 76-81 GHz Eagle RaceRunner Ultra based on NXP MR3003 / S32R274 chipset. We use a corner reflector as reference. Both sensors, in addition to the measured scenario, are shown in Fig. \ref{fig:Scenario}. The FMCW waveform parameters in both sensors are shown in table \ref{tab:Scenario}. As a target we use a trihedral corner reflector of estimated radar cross section of 6 square meters at the operating band.  

\section{Results and discussion}

To demonstrate the validity of our approach we choose a heavily interfered measurement, with the target buried below the signal-plus-interference level. As a preliminary step, we apply a digital high-pass filter to the sampled signal to reduce the effect of cross-talk between elements, as it is several orders of magnitude higher than the expected power from target reflections. Then, we apply the same filter to the hypotheses in our model, to avoid introducing a mismatch.

For the interference mitigation algorithm we assume no prior knowledge about the time of characteristics of the interference, other than it being a CW or FMCW signal. We use the same two step search applied to the synthetic measurements. For the first search, $M_{\kappa} = 20$ slope hypotheses and $M_{\tau} = 300$ time-shift hypotheses. For the second search we use $M_{\kappa} = 20$  and $M_{\tau} = 100$. The received signal has $N = 229$ samples. If we assume the possible models $A$ are stored in memory and do not need to be computed, the total execution time approximates the execution time of OMP, which takes around 0.07 seconds in a MATLAB CPU implementation.

\begin{figure}[t]
  \centering
  \includegraphics[width=1\linewidth]{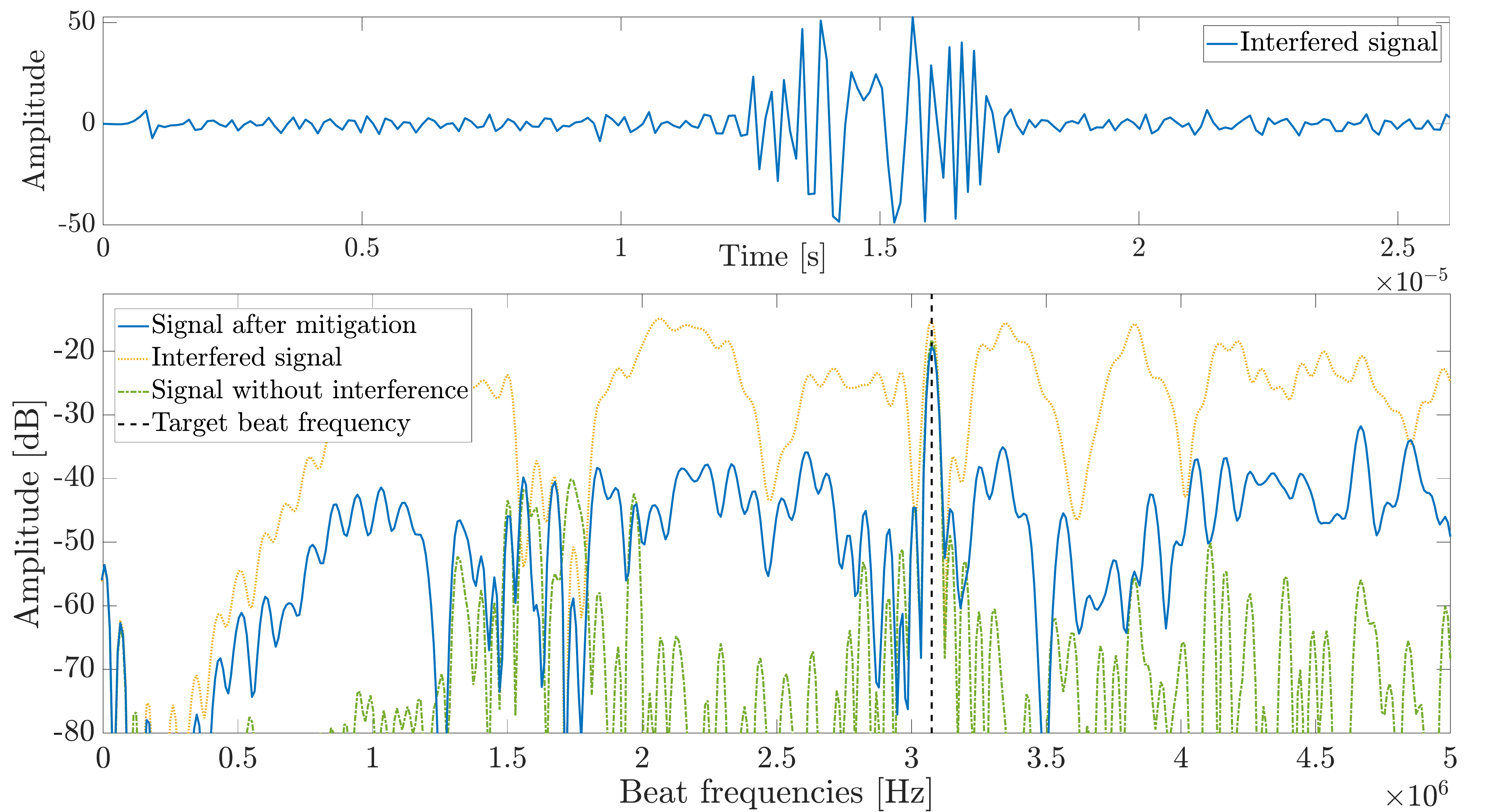}
  \caption[]{Interference mitigation applied to the interfered signal after high-pass filtering. We see a reduction of the SNIR level of up to 20 dB after mitigation, enough to recover our target.}
\label{fig:RangeEst}
\end{figure}

In Fig. \ref{fig:RangeEst}  we observe the range estimation of an interfered chirp. The target in the received, interfered signal can not be detected, as its power level is below the noise-plus-interference floor. After applying the proposed mitigation approach, the contribution of the interference is reduced and the target appears around 18 dB over the new signal-plus-interference floor. While a degradation in the SNIR can still be appreciated with the non-interfered measurement, the proposed method achieves an improvement of around 20 dB. The reduction in performance when compared to the synthetic cases is likely caused by hardware non-idealities in the receiver that are not captured by our model. Further research in this direction is likely to help increase the amount of interference removed.

\section{Conclusions}

In this paper a method to identify and mitigate interference in an automotive scenario is presented. The interference signal is reconstructed using a reduced chirplet basis and the interference contribution is subsequently removed by sparse reconstruction using the OMP algorithm. Results with synthetic and real data show that we can greatly diminish the effect of the interference, recovering targets that would otherwise be masked by it. OMP can be accelerated by hardware, making it a good choice for real time automotive applications.

\section*{Acknowledgment}

We thank our colleagues David Mateos-N\'{u}\~{n}ez and Carlos Moreno-Le\'on for fruitful discussions and their help during the measurement campaigns, and NXP Semiconductors for their assistance by providing one of the sensors for our tests.

\end{document}